\documentclass[reprint,
 amsmath,amssymb,
 aps,
]{revtex4-1}
\usepackage{color}
\usepackage[toc,page]{appendix}
\usepackage[english]{babel}
\usepackage{microtype}
\usepackage{braket}
\usepackage{graphicx}
\usepackage{dcolumn}
\usepackage{bm}
\usepackage[colorlinks]{hyperref}
\hypersetup{
    colorlinks=true,       
    linkcolor=blue,         
    citecolor=blue,        
    filecolor=blue,     
   urlcolor=blue         
}

\begin{document}

\title{Antiferromagnetic self-ordering of a Fermi gas in a ring cavity}
\author{Elvia Colella}
\email{elvia.colella@uibk.ac.at}
\affiliation{Institut f\"{u}r Theoretische Physik, Universit\"{a}t Innsbruck,
  Technikerstra{\ss}e 21a, A-6020 Innsbruck, Austria}
\author{Stefan Ostermann}
\affiliation{Institut f\"{u}r Theoretische Physik, Universit\"{a}t Innsbruck,
  Technikerstra{\ss}e 21a, A-6020 Innsbruck, Austria}
\author{Wolfgang Niedenzu}
\affiliation{Institut f\"{u}r Theoretische Physik, Universit\"{a}t Innsbruck,
  Technikerstra{\ss}e 21a, A-6020 Innsbruck, Austria}
\author{Farokh Mivehvar}
\affiliation{Institut f\"{u}r Theoretische Physik, Universit\"{a}t Innsbruck,
  Technikerstra{\ss}e 21a, A-6020 Innsbruck, Austria}
\author{Helmut Ritsch}
\email{helmut.ritsch@uibk.ac.at}
\affiliation{Institut f\"{u}r Theoretische Physik, Universit\"{a}t Innsbruck,
  Technikerstra{\ss}e 21a, A-6020 Innsbruck, Austria}
\begin{abstract}

We explore the density and spin self-ordering of driven spin-$1/2$ collisionless fermionic atoms coupled to the electromagnetic fields of a ring resonator. The two spin states are two-photon Raman-coupled via a pair of degenerate counterpropagating cavity modes and two transverse pump fields. In this one-dimensional configuration the coupled atom-field system possesses a continuous $U(1)$ translational symmetry and a discrete $\mathbf{Z}_2$ spin inversion symmetry. At half filling for sufficiently strong pump strengths, the combined $U(1)\times \mathbf{Z}_2$ symmetry is spontaneously broken at the onset of a superradiant phase transition to a state with self-ordered density and spin structures. We predominately find an antiferromagnetic lattice order at the cavity wavelength. The self-ordered states exhibit unexpected positive momentum pair correlations between fermions with opposite spin. These strong cavity-mediated correlations vanish at higher pump strength.
\end{abstract}

\maketitle                        

\section{\label{sec:level}Introduction}
Laser manipulation and control of cold atomic gases has recently seen spectacular advances of experimental technology~\cite{bloch2012quantum} as well as theoretical modelling~\cite{cohen2011advances, lewenstein2012ultracold,goldman2016topological,zhai2015degenerate}. In combination with state-of-the-art cavity technology it is now possible to routinely explore the dynamics of degenerate quantum gases in high-$Q$ optical cavities~\cite{kruse2004cold,ritsch2013cold,wang2012spin,mekhov2012quantum}. Numerous intriguing quantum phenomena ranging from spontaneous crystallization to supersolidity or non-trivial magnetic ordering have been predicted and experimentally observed~\cite{leonard2017supersolid,kohler2018negative,davis2018photon,kroeze2018spinor,landini2018formation}. Although so far experiments have been limited to bosonic atoms with only one or two internal states contributing to the dynamics, experiments using fermionic gases are well in reach and realizable with current technology. 
\par 
The opto-mechanical coupling of the atom and the cavity fields allows for dynamical trapping of the atoms and even cavity cooling of the gas towards quantum degeneracy~\cite{wolke2012cavity,sandner2013subrecoil}. As a decisive new feature, cavity modes can be designed to introduce tailored long-range interactions~\cite{vaidya2018tunable} and dynamic gauge fields for the ultracold atoms~\cite{Mivehvar2014SOC-B,ballantine2017meissner,halati2017cavity}. Thus, atom-cavity systems have proven to be a versatile basis for quantum simulations of exotic phases~\cite{leonard2017supersolid,leonard2017monitoring,georges2018light} with a wealth of further theoretical proposals still open for implementation ~\cite{mivehvar2017disorder,ostermann2018cavity}. Generalizations to many field modes and laser frequencies should allow the implementation of fully connected quantum annealing~\cite{torggler2017quantum,vaidya2018tunable}. 
\par
\begin{figure}[b!]
\centering
 \includegraphics[width=0.5\textwidth]{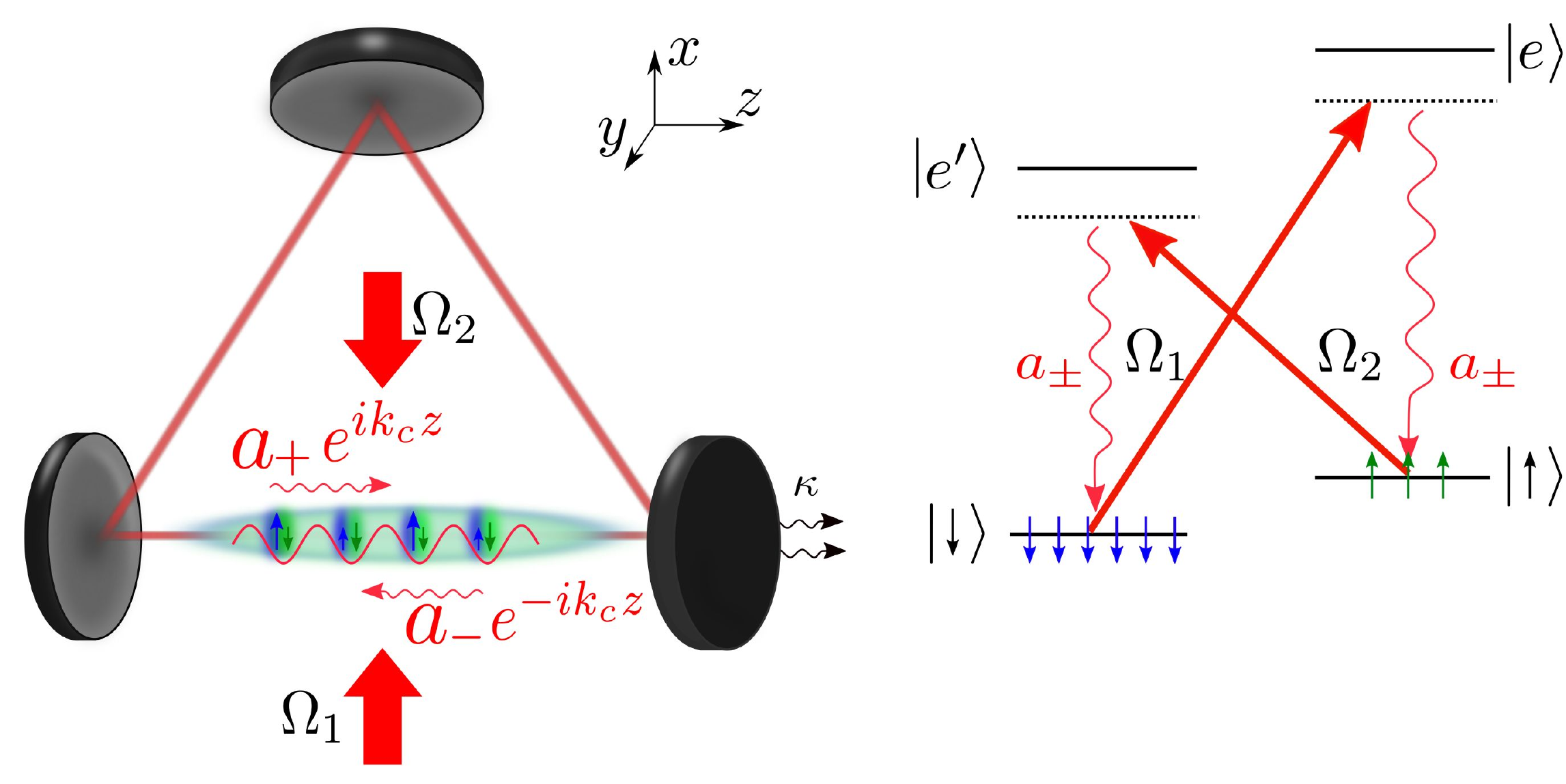}
 \caption{Schematic view of the system and the atomic-field coupling. The transitions between two atomic ground states $\{\ket{\uparrow},\ket{\downarrow}\}$ and two excited states $\{\ket{e'},\ket{e}\}$ are induced via two far red-detuned lasers with Rabi frequencies $\Omega_{1,2}$ and two cavity field modes with coupling strengths $g_0 e^{\pm i k_cz }$.}
 \label{scheme}
 \end{figure}

 
With the prediction of new intriguing phenomena such as Umklapp superradiance~\cite{piazza2014umklapp,chen2014superradiance}, topologically protected edge states ~\cite{mivehvar2017superradiant,sheikhan2016cavity}, superconducting pairing~\cite{colella2018quantum,sheikhan2018cavity}, artificial dynamic gauge fields~\cite{halati2017cavity}, unconventional momentum correlations and quantum phases in multiple dimensions~\cite{fan2018magnetic,feng2017quantum,sandner2015self,chen2015superradiant}, implementations of fermionic systems coupled to cavity fields have gained more attention recently. In the present article, we propose the realization of  density and spin self-ordering for a transversely driven multi-level Fermi gas coupled to a pair of counterpropagating degenerate modes of a ring cavity as depicted in Fig.~\ref{scheme}~\cite{kruse2003observation,slama2007cavity,slama2007superradiant,bux2013control,schmidt2014dynamical,culver2016collective,naik2018bose}. The multi-level atomic structure allows to implement spinor states \cite{kroeze2018spinor}, while the cavity geometry guarantees a continuous translational symmetry \cite{schuster2018pinning,wolf2018observation}. The dynamical coupling between the light fields and the atomic states induces a transversal spin-wave texture of antiferromagnetic nature~\cite{Mivehvar2018toolbox}. We show that the common interaction of the atoms with the cavity fields results in the build up of unexpected positive momentum correlations between the atoms. 

This paper is organized as follows. In Sec.~\ref{sec:level1} we introduce the model. In Sec.~\ref{sec:level2} we derive the mean-field coupled equations of motion and discuss our numerical approach. We then present the main numerical results and describe the phase diagram of the system in Sec.~\ref{sec:level3}A. The superradiance transition threshold is analytically obtained in Sec.~\ref{sec:level3}B and the nature of the density and spin self-organized states is further discussed in Sec.~\ref{sec:level3}C. Section~\ref{sec:level4} is devoted to analyse the photon-induced momentum correlations between the atoms. Concluding remarks are presented in Sec.~\ref{sec:level5}.

\section{\label{sec:level1}Model}
Consider an ensemble of transversely-driven ultracold, fermionic four-level atoms coupled to two modes of a ring cavity as shown in Fig.~\ref{scheme}. The two atomic ground states $\{\ket{\uparrow},\ket{\downarrow}\}$ with energies $\{\hbar\omega_\uparrow,\hbar\omega_\downarrow=0\}$ are coupled to two excited states $\{\ket{e},\ket{e'}\}$ with energies $\{\hbar \omega_e,\hbar \omega_{e'}\}$ through the interaction with the cavity fields and two external classical pump fields. The energy difference between the two ground states can be tuned by an external longitudinal magnetic field $B_z$. The atoms are assumed to be strongly confined in the transverse directions, therefore, their motion is restricted along the cavity axis. The ring cavity supports a pair of degenerate counterpropagating modes $\hat a_{\pm}e^{\pm i k_c z}$ with the same linear polarization and frequency $\omega_c$, and opposite wave-numbers $\pm k_c=\pm2\pi/\lambda_c$. Here, $\hat a_{+}$ ($\hat a_-$) is a bosonic field operator annihilating a photon in the forward (backward) propagating cavity mode. The atoms are pumped from the side by two lasers with frequencies $\{\omega_{p1},\omega_{p2}\}$ and opposite circular polarizations, where we have chosen the quantization axis along the cavity axis. The two classical laser fields with Rabi frequencies $\Omega_1$ and $\Omega_2$ drive the transitions $\ket{\downarrow}\leftrightarrow\ket{e}$ and $\ket{\uparrow}\leftrightarrow\ket{e'}$, respectively. Without loss of generality we take both Rabi frequencies to be real. The cavity fields $\hat a_{\pm}$ couple both transitions $\ket{e'}\leftrightarrow\ket{\downarrow}$ and  $\ket{e}\leftrightarrow\ket{\uparrow}$, with coupling strengths $g_{0}e^{\pm i k_c z}$.
\par
In the limit where the pumps and cavity fields are far detuned from the atomic excited states $\ket{e}$ and $\ket{e'}$, only virtual excitations are created and the system can be effectively described as a spin-$1/2$ system $\{\ket{\downarrow},\ket{\uparrow}\}$. As discussed in~\ref{app:1}, this system is described by the effective time-independent Hamiltonian 
 \begin{align}
 \label{ham}
 H&=\sum_{\sigma\in\{\uparrow,\downarrow\}} \int dz \ \hat\Psi^\dagger_\sigma(z)\Big[-\frac{\hbar^2}{2m}\frac{d^2}{dz^2}+\hbar \delta_\sigma +\hat{U}_{\sigma}(z) \Big]\hat\Psi_\sigma(z)\nonumber\\&\quad+ \int dz \ \hbar \hat{\eta}_R (z) \ \Big[\hat\Psi^\dagger_{\uparrow}(z)\hat\Psi_\downarrow(z)
 +\hat \Psi^\dagger_{\downarrow}(z)\hat\Psi_\uparrow(z) \Big]\nonumber\\
 &\quad-\hbar \Delta_c (\hat a^\dagger_{+}\hat a_{+}+\hat a^\dagger_{-}\hat a_{-})
 , 
\end{align}
 where $\hat\Psi_\sigma(z)$ are fermionic field operators fulfilling the anti-commutation relation $\{\hat\Psi_\sigma(z),\hat\Psi_{\sigma'}^\dagger(z')\}=\delta(z-z')\delta_{\sigma,\sigma'}$.
 The effective detunings between the two spin states and the pump fields are denoted as $ \delta_{\downarrow}=0$ and $ \delta_{\uparrow}=\omega_{\uparrow}+B_z-(\omega_{p2}-\omega_{p1})/2$, respectively. That is, $\hbar\delta\equiv\hbar(\delta_\uparrow-\delta_\downarrow)$ defines the effective energy splitting between the two spin states.

In this model photons interact with atoms via two fundamental mechanisms. The scattering of photons by the atoms between the two cavity modes $\hat a_{\pm}$ induces a potential with $\lambda_c/2$ periodicity, 
 \begin{equation}
\hat{U}_\sigma(z)=U_{0\sigma} (\hat a^\dagger_{+}\hat a_{+}+\hat a^\dagger _{-}\hat a_{-}+e^{-i2k_cz} \hat a^\dagger _+ \hat a_-+e^{i2k_cz} \hat a^\dagger _- \hat a_+),
\label{usigma}
 \end{equation}
with  $U_{0\uparrow}= \hbar g_0^2/\Delta_e$ and $U_{0\downarrow}=\hbar g_0^2/\Delta_{e'} $, where $\Delta_e=(\omega_{p1}+\omega_{p2})/2-\omega_e$ and $\Delta_{e'}=\omega_{p1}-\omega_{e'}$. On the other hand,  scattering of photons from the pumps into the cavity modes by the atoms results in spin flipping processes with $\pm \hbar k_c$ momentum kicks to the atoms described by the $\lambda_c$-periodic Raman coupling term
\begin{equation}
\hat{\eta}_R(z)=\eta (\hat a_+ e^{ik_cz}+\hat a_-e^{-ik_cz}+\hat a_+^\dagger e^{-ik_cz}+\hat a_-^\dagger e^{ik_cz}).
\label{etar}
\end{equation}
We have considered the balanced Raman coupling configuration $\eta\equiv \Omega_1 g_0/\Delta_{e}=\Omega_2 g_0/\Delta_{e'}$, where $\eta$ is the two-photon Rabi frequency. 

The last line in the Hamiltonian \eqref{ham} represents the energy contribution of the two cavity modes $\hat a_\pm$, where the cavity detuning is $\Delta_c=(\omega_{p1}+\omega_{p2})/2-\omega_c$. Cavity losses will be phenomenologically included in the equations of motion for the field operators $\hat a_\pm$ via the cavity decay rate $\kappa$ \cite{wallsbook}. Note that contact two-body interactions between atoms are assumed to be negligible throughout this work.

The processes acting on the spin and the density degrees of freedom are characterized by competing periodicities. The periodic potentials $\hat{U}_\sigma(z)$ [Eq.~\eqref{usigma}] favour the organization of the atoms in a $\lambda_c/2$-periodic structure. By contrast, the position dependent Raman coupling $\hat{\eta}_R(z)$ [Eq.~\eqref{etar}] favours a spin texture with $\lambda_c$ periodicity. Therefore, the Raman coupling term defines the periodicity of the Hamiltonian and the size of the Brillouin zone $[-k_c/2,k_c/2]$.
\par
The Hamiltonian \eqref{ham}  is invariant under the parity transformation of the photonic operators $\hat a_{\pm}\rightarrow - \hat a_{\pm}$ and the $\pi$-rotation of the local transverse spin of the system $\hat S_x(z)\rightarrow -\hat S_x(z)$ and $\hat S_y(z)\rightarrow - \hat S_y(z)$. Here, the total local spin operator is defined as
\begin{equation}
\mathbf{\hat{S}}(z)=
      \left(\hat \Psi^\dagger_\uparrow(z),\hat \Psi^\dagger_\downarrow(z)\right)
      \boldsymbol{\tau}
      \begin{pmatrix}
      \hat \Psi_\uparrow(z) \\
      \hat \Psi_\downarrow(z)
     \end{pmatrix} ,
     \label{svector}
\end{equation}
where $\boldsymbol{\tau}=(\tau_x,\tau_y,\tau_z)$ is the vector of the Pauli matrices. The combination of parity and spin inversion yields a discrete $\mathbf{Z}_2$ symmetry. In addition, the Hamiltonian \eqref{ham} is invariant under the simultaneous transformations  $z\rightarrow z+\Delta z$ and $\hat a_{\pm} \rightarrow \hat a_{\pm} e^{\mp ik_c \Delta z} $, yielding a continuous $U(1)$ symmetry. Any arbitrary displacement of the position of the atom can be compensated by a phase shift of the photonic operators. This continuous $U(1)$ symmetry is a specific character of the ring cavity geometry~\cite{mivehvar2018driven}. In fact, in a ring cavity the intensity maxima of the cavity fields can sit at any position on the cavity axis,  realizing a continuous translational symmetry.  This is in sharp contrast to linear cavities where the cavity fields must have a node on the mirrors to satisfy the boundary conditions, giving rise to a discrete $\mathbf{Z}_2$ symmetry~\cite{mivehvar2017superradiant}. 

Therefore, the Hamiltonian \eqref{ham} possesses  a $ U(1)\times \mathbf{Z}_2$ symmetry, which is spontaneously broken at the onset of the superradiant phase transition with the emergence of a self-organized density and spin texture, as will be shown in the following. In contrast to single component quantum gases~\cite{baumann2010dicke,baumann2011exploring}, where the phase transition is driven by a density order parameter, here the spin self-organization plays the fundamental role in the superradiant process~\cite{mivehvar2017disorder,yu2017topological,pan2015topological}. In other words, the cavity modes can only be populated for a non-vanishing spin order parameter.

\section{\label{sec:level2}Self-consistent mean-field method}
In order to determine the steady state of the system we employ a self-consistent mean-field method. The atomic state is dynamically coupled to the cavity-photon dynamics, see~\ref{app:2}. The large cavity detuning $|\Delta_c|$ and cavity linewidth $2\kappa$ dictate a fast dynamical evolution of the cavity fields, which at each moment adiabatically follows the atomic state~\cite{domokos2001semiclassical}. On the other hand, the fermionic dynamics evolve in the self-consistent potentials and the Raman field created by the cavity modes (and pump lasers). At a given value of the cavity fields $\langle \hat a_\pm\rangle =\alpha_\pm$, the atomic dynamics can be described by a single-particle Hamiltonian. Upon making a Bloch ansatz for the single-particle wave function
$\psi_{nq\sigma}(z)=e^{i qz} u_{nq\sigma}(z)$~\cite{kittel1965quantum}, the atomic field operators can be expanded in the basis of the Bloch functions,
\begin{equation}
   \hat \Psi_\sigma(z)=\sum_{n,q} \psi_{nq\sigma}(z)\hat c_{nq}.
\end{equation}
Here $\hat c_{nq}$ is a fermionic operator which annihilates a particle in the $n$th band with quasi-momentum $q$ and $u_{nq\sigma}(z)$ are $\lambda_c$-periodic functions. Therefore, the single particle problem is solved by diagonalizing the Hamiltonian within one unit cell $[0,\lambda_c=2\pi/k_c]$ with periodic boundary conditions. We aim to determine the eigenvalues $\epsilon_{nq}$ of the coupled Schr\"odinger equations for the functions $u_{nq\uparrow}(z)$ and $u_{nq\downarrow}(z)$,
\begin{subequations}
\begin{align}
\Big[\frac{\hbar^2}{2m}\Big(i\frac{d}{dz}-q\Big)^2+\hbar\delta_\uparrow+U_\uparrow(z)\Big]u_{nq\uparrow}(z)&+\hbar\eta_R(z)u_{nq\downarrow}(z)\\\nonumber&=\epsilon_{nq}u_{nq\uparrow}(z),\\
\Big[\frac{\hbar^2}{2m}\Big(i\frac{d}{dz}-q\Big)^2
+\hbar\delta_\downarrow+U_{\downarrow}(z)\Big]u_{nq\downarrow}(z)&+\hbar\eta_R(z)u_{nq\uparrow}(z)\\\nonumber&=\epsilon_{nq}u_{nq\downarrow}(z),
\end{align}
\label{seteqtns}
\end{subequations}
where the quasi-momentum $q$ lies in the first Brillouin zone, $q\in[-k_c/2,k_c/2]$.  
\par
The chemical potential $\mu$ of the system has to be determined self-consistently by fixing the total number of particles
\begin{equation}
N=\sum_\sigma \int_{0}^{L} dz \ n_\sigma(z),
\end{equation}
 where 
\begin{equation}
n_\sigma(z)=\sum_{nq} |u_{nq\sigma}(z)|^2n_F(\epsilon_{nq})
\end{equation}
is the local atomic density in the $\sigma$-spin state with $n_F(\epsilon)=1/[1+e^{(\epsilon-\mu)/k_BT}]$ being the Fermi distribution. We assume thermal equilibrium between the two spin states and therefore use the same chemical potential $\mu$ for both states throughout the calculation.

Equations \eqref{seteqtns} are solved  in a self-consistent way in combination with the stationary values of the cavity fields $\alpha_\pm$. As discussed in~\ref{app:2}, the stationary field amplitudes are given by
\begin{subequations}
\begin{align}
\alpha_+ &= \frac{2\eta(\tilde{\Delta}_c+i\kappa)}{(\tilde{\Delta}_c+i\kappa)^2-U_0^2|\mathcal{N}_{2k_c}|^2}\Big(\Theta^*+\frac{U_0\mathcal{N}_{2k_c}^*}{\tilde{\Delta}_c+i\kappa}\Theta \Big),\\
\alpha_-& = \frac{2\eta(\tilde{\Delta}_c+i\kappa)}{(\tilde{\Delta}_c+i\kappa)^2-U_0^2|\mathcal{N}_{2k_c}|^2}\Big(\Theta+\frac{U_0\mathcal{N}_{2k_c}}{\tilde{\Delta}_c+i\kappa}\Theta^* \Big).
\end{align}
\label{eqalphas}
\end{subequations}
Here we have defined the effective shifted cavity detuning $\tilde\Delta_c=\Delta_c-U_0N$ and the atomic averages
\begin{equation}
\quad\mathcal{N}_{2k_c}= \int dz e^{2ik_c z}n(z),
\label{cdw}
\end{equation}
and
\begin{equation}
\Theta=\int dz e^{ik_cz} S_x(z),
\label{sdw}
\end{equation}
where $\Theta$ is the spin order parameter driving the superradiant phase transition. Indeed, a non-vanishing $\Theta$ is required for non-zero cavity fields in Eqs.~\eqref{eqalphas}. Here, 
\begin{equation}
    n(z)=n_\uparrow(z)+n_\downarrow(z)\label{densitynz}
\end{equation} is the total atomic density and 
\begin{equation}
S_x(z)=\frac{1}{2}\sum_{q,n} \Big[u_{nq\uparrow}^*(z)u_{nq\downarrow}(z)+u_{nq\downarrow}^*(z)u_{nq\uparrow}(z)\Big]n_F(\epsilon_{qn})    
\label{Sx}
\end{equation}
is the average local spin component in the $x$-direction. These atomic averages can also be interpreted as the probabilities of photon-atom scattering processes. In fact, 
$ \mathcal{N}_{2k_c}$ is the probability that an atom absorbs a photon and then re-emits it in the opposite direction receiving a $2\hbar k_c$ momentum kick without changing its internal state. On the other hand, $\Theta$ is the probability of scattering a photon from a pump laser into a cavity mode where the atom changes both its internal and external states, with a momentum exchange of  $\hbar k_c$. 

\begin{figure*}[t]
 \includegraphics[width=1
\textwidth]{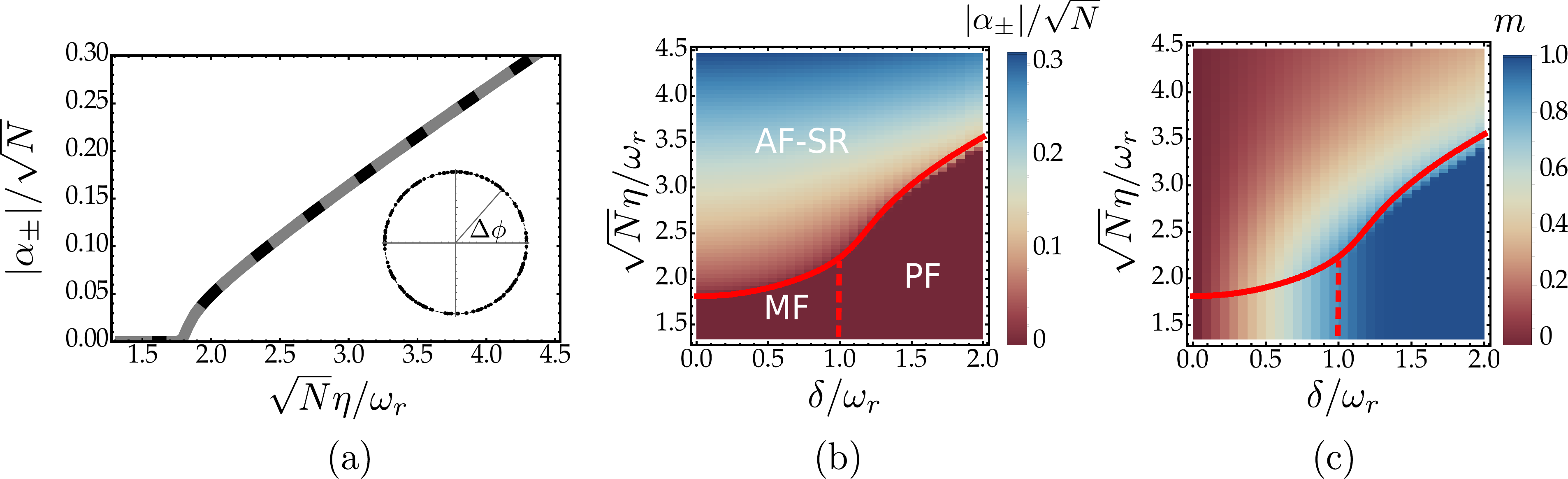}
 \caption{(a) Intra-cavity amplitudes of the two modes $\alpha_{+}$ $(\alpha_{-})$ as a function of the rescaled pump strength $\sqrt{N}\eta/\omega_r$ for the effective energy splitting $\delta=0$ shown as grey (dashed black) curves. Inset: relative phase $\Delta\phi=\phi_+-\phi_-$ of the two modes for 200 realizations demonstrating the $U(1)$ symmetry breaking. Parameters: $\{\delta,\sqrt{N} \eta \}=\{0,2.7\}\omega_r$. (b)  Phase diagram of the system. The color encodes the amplitudes $|\alpha_\pm|/\sqrt{N}$ of the fields. Below threshold, the system can be either a mixed Fermi (MF) gas or a polarized Fermi (PF) gas, separated by a dashed red line. Above threshold, we find a superradiant state with antiferromagnetic character (AF-SR). The red solid line is the analytical result \eqref{critpump} for the critical pump strength. (c) Global longitudinal magnetization \eqref{magn} of the system. Cavity parameters: $\{\Delta_c,\kappa,U_0 N \}=\{-20,10,-8\}\omega_r$.}
 \label{phased}
 \end{figure*}

\section{\label{sec:level3}Superradiant phase transition}
In the following, we characterize the phase diagram of a Fermi gas at fixed density, $k_F/k_c=1/2$, i.e., half filling, with $k_F$ being the Fermi momentum. The single-particle Hamiltonian is diagonalized within one unit cell with periodic boundary condition and fifty  quasi-momenta $q$ (equivalent to a lattice of $N_c=50$ sites) at finite temperature $k_BT=0.05 \hbar\omega_r=0.2 k_B T_F$. Here, $T_F$ is the Fermi temperature, defined as $k_B T_F =\hbar^2 k_F^2/2m$, and $\omega_r=\hbar k_c^2/2m$ is the recoil frequency. The cavity is red detuned, $\Delta_c=-20\omega_r$. The presence of the Fermi gas induces a shift of the cavity frequency proportional to the refractive index, $U_0N=-8\omega_r$. For a fixed value of the refractive index, the effective shifted cavity detuning $\tilde\Delta_c=\Delta_c-U_0N=-12\omega_r$ is still in the red detuned regime. The cavity linewidth is chosen as $\kappa=10\omega_r$.       

\subsection{Phase diagram}
In Fig.~\ref{phased}(a) we show the amplitudes $|\alpha_\pm|/\sqrt{N}$  of the two cavity fields as a function of the rescaled pump strength $\sqrt{N}\eta/\omega_r$ for the degenerate case, where the effective level splitting vanishes, $\delta=0$. In our mean-field picture, in the superradiant phase the cavity fields are coherent states, $\langle \hat{a}_\pm \rangle=\alpha_\pm= |\alpha_\pm |e^{i \phi_\pm}$. The two modes are symmetrically coupled to the two atomic transitions and hence are equally populated. Their amplitudes grow monotonically across the transition point, hinting to the occurrence of a second order phase transition. Above threshold the field amplitudes scale as $\sim \eta ^{3/2}$ which differs from the $\sim \eta ^{1/2}$ power law exponent found in conventional self-organization in standing-wave cavities \cite{asboth2005self,niedenzu2011kinetic}. The phase difference of the two modes,  $\Delta\phi=\phi_+- \phi_-$, can acquire any value between 0 and $2\pi$ as shown in the inset of Fig.~\ref{phased}(a), where the relative phase of the two fields is shown for 200 realizations at fixed parameters. The relative phase uniformly distributes on a circle, demonstrating the continuous $U(1)$ symmetry breaking. This implies that the minima of the optical potential generated by the interference of the two cavity modes can be located anywhere within one unit cell. The translational symmetry is therefore connected with the relative phase of the two modes and the system possesses a full $U(1)$ symmetry. 
\par
In Fig.~\ref{phased}(b) the amplitudes of the two modes are shown as a function of the effective level splitting $\delta/\omega_r$ and the rescaled pump strength $\sqrt{N}\eta/\omega_r$. At each fixed $\delta$ the transition to the superradiant state, indicated by the solid red line, is of second order. Note that the critical threshold grows with increasing atomic energy spacing $\delta$. In Fig.~\ref{phased}(c) we show the global magnetization 
\begin{equation}
m=\frac{N_\uparrow-N_\downarrow}{N},
\label{magn}
\end{equation} 
of the atomic gas in the same parameter space, where $N_\sigma=\int_0^L dzn_\sigma(z)$. Below the superradiant transition threshold the Fermi gas is in a trivial phase where the population imbalance is governed by the energy difference $\delta$ between the two spin states. In fact, this parameter acts as an effective longitudinal magnetic field, orienting the spin of the particles in its direction. For $\delta=0$ the system is not magnetized, $m=0$. With increasing $\delta$, an increasing amount of atoms align with the effective magnetic field and the system is an incoherent mixed Fermi (MF) gas, $0<m<1$. The mixed phase is arising from the incoherent superpositions of the atoms in the two spin states. It is a direct result of the thermalization of the atoms in the Zeeman sub-levels. The studied system can thermalize via two-body contact interactions, incoherent decay or Raman transitions induced by thermal photons or vacuum fluctuations. For $\delta>\omega_r$ all particles are aligned in the same direction and the system becomes a polarized Fermi (PF) gas with $m=1$. In the superradiant regime, the magnetization of the system gradually decreases, evolving towards an ordered state of antiferromagnetic character with zero magnetization $m=0$, which exhibits superradiant photon scattering (AF-SR).

\subsection{Transition threshold}
Due to the continuous change of the order parameters $\alpha_\pm$ across the critical point, the superradiant phase transitions can be described within the framework of the Landau theory of second-order phase transitions \cite{Landau1937onthetheoryofphasetransition}. The transition threshold can be obtained by expanding the free energy in powers of the order parameter and requiring that the coefficient of the second order term vanishes at the critical point. Integrating out fermionic degrees of freedom \cite{yu2017topological,pan2015topological,chen2014superradiance}, the free energy as a functional of the cavity-field order parameters $\alpha_\pm$ is expressed as 
\begin{equation}
  F[\alpha_\pm^*,\alpha_\pm]=-\tilde{\Delta}_c \left(|\alpha_-|^2+|\alpha_+|^2\right)-\eta^2 N\chi_m |\alpha_+^*+\alpha_-|^2.
\end{equation}
Here only terms up to second order in $\alpha_\pm$ are retained and 
\begin{equation}
    \chi_m=\sum_{k} \frac{n_F(\epsilon_{ k+k_c\uparrow})-n_F(\epsilon_{ k\downarrow})}{\epsilon_{ k\downarrow}-\epsilon_{ k+k_c\uparrow}}
\end{equation}
is the magnetic susceptibility of the Fermi gas, where $\epsilon_{k\uparrow}=\hbar^2k^2/2m-\mu+\delta$ and  $\epsilon_{k\downarrow}=\hbar^2k^2/2m-\mu$ are the bare energies of the two fermionic spins before the transition.\par
We express the free energy as a functional of the atomic order parameter $\Theta$ using Eq.~\eqref{eqalphas}. In a first-order approximation we neglect the contribution of the mixing of the two cavity fields due to scattering processes which stem from the optical potentials $U_\sigma(z)$. This approximation is well justified in our parameter regime $U_0|\mathcal{N}_{2k_c}|/\tilde{\Delta}_c\ll U_0N/\tilde{\Delta}_c\sim 1 $ at the onset of the phase transition.
The quadratic free energy in terms of the atomic order parameter therefore takes the form 
\begin{equation}
    F[\Theta^*,\Theta]\sim \Big(1-\frac{\eta^2}{\eta_c^2}\Big)|\Theta|^2,
\end{equation}
where
\begin{equation}
\sqrt N\eta_c=\sqrt{\frac{\tilde{\Delta}_c^2+\kappa^2}{2\tilde{\Delta}_c\chi_m}}
\label{critpump}
\end{equation}
is the critical pump strength, which depends on the cavity parameters $\tilde{\Delta}_c$ and $\kappa$ and on the magnetic susceptibility $\chi_m$. At zero temperature and for the degenerate case $\delta=0$, the divergence of $\chi_m$ leads to a strong suppression of the transition threshold. In fact, the scattering of a photon from the pump into the cavity by an atom results in a $k_c=2k_F$ momentum transfer, causing the atom to scatter from one side of the Fermi surface to the other. This process requires nearly no energy cost and leads to a vanishing critical pump strength at $T=0$. In order for this to occur, the nesting condition $k_c=2k_F$ must be satisfied, in analogy to polarized fermions in linear cavities \cite{piazza2014umklapp,chen2014superradiance,mivehvar2017superradiant,sandner2015self}. A finite temperature washes out the divergence of the magnetic susceptibility, resulting in a finite, although small threshold. In analogy to the one component case, superradiance should be robust against thermal fluctuations. At high temperature, the atomic system can be described as a classical gas following the Boltzmann statistics. An expansion of Eq.~\eqref{critpump} for high temperatures reveals a $T^{1/2}$ scaling of the transition threshold. For sufficiently high pump strengths, this should still allow to observe the transition to the magnetic state. In addition, the presence of a finite energy splitting $\delta$ shifts the nesting wave-vector, resulting in an increase of the critical threshold in comparison to $\delta=0$, where the nesting condition perfectly holds. Similarly deviations from the nesting condition, due to trapping inhomogeneities or the incommensurability between the cavity wave-vector and the Fermi momentum, increase the critical threshold but preserve superradiance, as it is shown in~\ref{app:3}. The critical threshold \eqref{critpump} is shown in Figs.~\ref{phased}(b) and~\ref{phased}(c) as a red solid curve separating the superradiant regime from the trivial phase. The analytic threshold is consistent with the numerical result. Small deviations from the theoretical prediction for big level spacing $\delta$ can be attributed to neglected terms scaling with $U_0|\mathcal{N}_{2k_c}|/\tilde{\Delta}_c$ in the analytic calculation.\\

\begin{figure}[t]
\centering \includegraphics[width=0.5\textwidth]{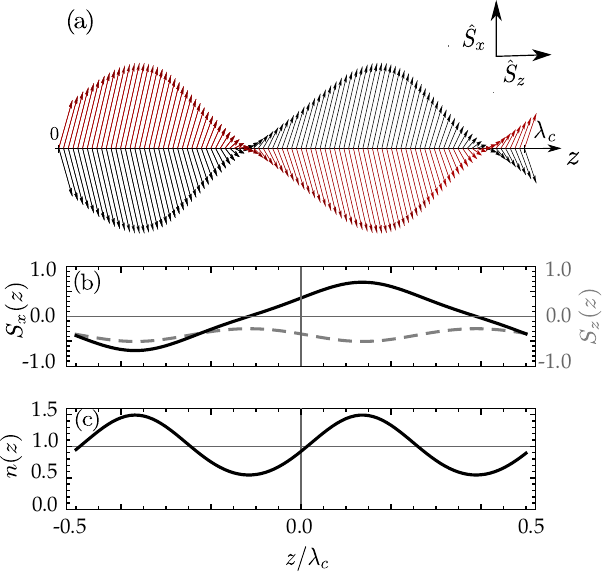}
 \caption{Atomic structure above threshold, $\eta>\eta_c$. (a) Spin texture within one unit cell $[0,\lambda_c]$ in the $\{S_x,S_z\}$ plane: note that $S_y(z)=0$. The two possible spin textures are shown in black and red for a given density configuration, exhibiting the $\textbf{Z}_2 $ symmetry of the system as described in the main text. (b) Individual local spin components $S_x(z)$ and $S_z(z)$, and (c) total atomic density distribution  $n(z)=n_\uparrow(z)+n_\downarrow(z)$. Parameters: $\sqrt{N}\eta=2.7 \omega_r$ and $\delta=0.6\omega_r$. Other parameters as in Fig.~\ref{phased}.}
 \label{spintex}
 \end{figure} 
  \begin{figure}[t]
\centering
 \includegraphics[width=0.5\textwidth]{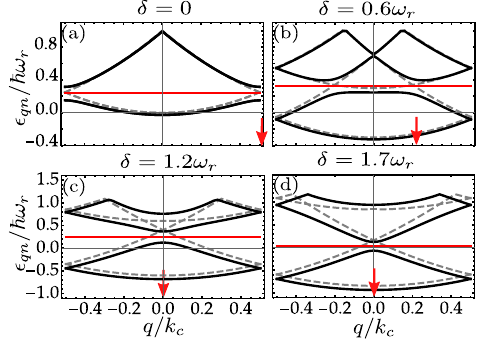}
  \caption{Atomic band structure both below threshold (dashed gray) and above the ordering transition (black solid). The position of the chemical potential (red line) indicates a metal-insulator transition. The red arrows show the quasimomentum $q$ at which the insulating gap opens. The energy splitting is (a) $\delta=0\omega_r$, (b) $\delta=0.6\omega_r$, (c) $\delta=1.2\omega_r$ and (d) $\delta=1.7\omega_r$.  }
 \label{BS}
 \end{figure}

\subsection{Self-organization}
In Fig.~\ref{spintex} we illustrate the structure of the Fermi gas in the superradiant phase by analysing the behaviour of the local density and spin of the system. Figure~\ref{spintex}(a) illustrates the local spin vector $\langle \mathbf{\hat S}\rangle$ [cf. Eq.~\eqref{svector}] as a function of the position along the cavity axis within one unit cell. The individual $S_x(z)$ and $S_z(z)$ components are shown in Fig.~\ref{spintex}(b). Note that the spin of the system always lies in the $\{S_x,S_z\}$ plane, i.e., $S_y(z)=0$. A non-vanishing $S_z(z)$ component is induced by the effective energy spacing $\delta$ acting as an effective longitudinal magnetic field. In addition, the Raman coupling acts as a transversal magnetic field in the $x$-direction, which adiabatically drives the $S_x(z)$ spin component~\cite{Mivehvar2018toolbox}. The optical potential $U_\sigma(z)$ favors a $\lambda_c/2$-periodic density pattern, as shown in Fig.~\ref{spintex}(c).
\par
For intermediate pump strengths, the system is weakly self-organized and the spin texture is characterized by the presence of a transversal $\lambda_c$-periodic spin wave in the $x$-direction, $S_x(z)$, and a longitudinal $\lambda_c/2$-periodic spin wave in the $z$-direction, $S_z(z)$, see Fig.~\ref{spintex}(b). The $S_x(z)$ component spontaneously emerges from the interference of two counterpropagating photon-induced spin waves. The phenomenon has common features with itinerant antiferromagnetism in Chromium and Chromium alloys~\cite{fawcett1994spin,fawcett1988spin, overhauser1959new}. At very high pump strengths, the system is strongly self-organized and gradually evolves toward an antiferromagnetic state in a more conventional sense. The corresponding state arises from the freezing of the spin degrees of freedom, which would result in a reduced entropy per particle in the self-organized state. The global magnetization drops to zero (see Fig.~\ref{phased}(c)) and the $S_z(z)$ component becomes negligible. In this regime the optical potentials $U_\sigma(z)$ localize the atoms, resulting in the emergence of a $\lambda_c$-periodic antiferromagnetic lattice order similar to bosonic atoms inside linear cavities~\cite{mivehvar2017disorder,Mivehvar2018toolbox}. 
\par
The position of the density peaks within the unit cell is arbitrarily chosen, indicating the spontaneous $U(1)$ symmetry breaking. However, for a given density configuration, the ground state is twofold denegerate. In Fig.~\ref{spintex}(a) we show the two degenerate spin textures (black and red). The $\mathbf{Z}_2$ symmetry breaking corresponds to the realization of one of the two possible spin textures.

\par
The spin and density structures  induce a metal-insulator transition which can be observed in the appearance of a gap in the atomic band structure. In Fig.~\ref{BS}, we show the band structure for different $\delta$ both in the superradiant regime (solid curves) and below threshold (dashed curves). The spin gap can open at any quasimomentum $q$, where the original $\ket \uparrow$ and $\ket \downarrow$ bands cross each other. In particular, at $\delta=0$, where the two states $\{\ket \uparrow,\ket  \downarrow\}$ are degenerate, the gap opens at the edges of the Brillouin zone $q=\pm k_c/2$. Increasing $\delta$ the gap opening gradually shifts toward zero until reaching the critical value $\delta=\omega_r$. In fact, by increasing the effective spin energy spacing the bands of the two states are gradually pushed apart, until for $\delta=\omega_r$ the $ \ket\downarrow$ states becomes energetically more favourable and the system becomes fully polarized. For $\delta>\omega_r$ the spin gap opens between higher bands. However, the presence of the self-consistent optical lattices $U_\sigma(z)$ favours the opening of a density gap at $q=0$, which preserves the insulating state.

  \begin{figure*}[t]
\centering
 \includegraphics[width=0.95\textwidth]{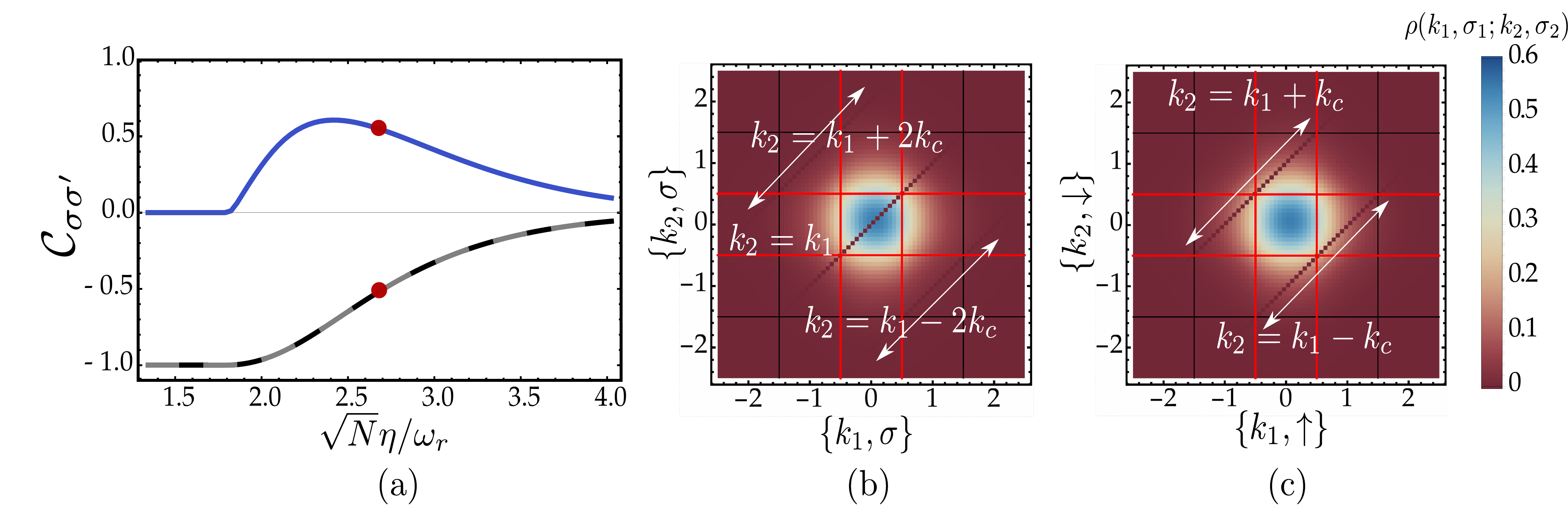}
 \caption{(a) Correlation coefficient for the same (gray and dashed black) and opposite spin directions (blue) as a function of the rescaled pump strength $\sqrt{N}\eta/\omega_r$ at the effective energy splitting $\delta=0$.  The red dot represents the pump value for which the two-body density matrix is shown. The two-body density matrix $\rho(k_1,\sigma_1;k_2,\sigma_2)$ in momentum space for $\delta=0$ and $\sqrt{N}\eta=2.7\omega_r$ for particles with (b) the same spin and (c) opposite spins.  The red lines show the position of the Fermi surface below transition, $k_F=1/2 k_c$ . The thermal background strongly deviates from the trivial phase, evolving from a perfect square below the threshold to a smooth circle above $\eta_c$. The interaction between cavity photons and fermions is responsible for the appearance of the off-diagonal dips at $k_1=k_2\pm k_c$ and $k_1=k_2\pm 2k_c$. The white arrows are guides to the eye for emphasising the off-diagonal dips.}
 \label{2bdm}
 \end{figure*} 

\section{\label{sec:level4}Momentum correlations}

Studies of momentum correlations between two particles in ring cavities revealed a strong coupling between the light fields and the atomic motion~\cite{gangl2000cold,niedenzu2012quantum}. In particular, while classical particles show a strong damping of the center-of-mass motion and anticorrelated momenta, quantum particles tend to correlate their motion~\cite{niedenzu2012quantum}. Quantum simulations of particles with Fermi or Bose statistics show momentum anticorrelation and correlation, respectively~\cite{sandner2013subrecoil}. The momentum correlation coefficient for two atoms in spin states $\sigma_1$ and $\sigma_2$ is defined as
\begin{equation}\label{eq_C}
  \mathcal{C}_{\sigma_1\sigma_2}=\frac{\langle k_1 k_2 \rangle_{\sigma_1\sigma_2}-\langle k_1\rangle_{\sigma_1}\langle k_2 \rangle_{\sigma_2}}{\Delta k_1 \Delta k_2},
\end{equation}
where $\mathcal{C}_{\sigma_1\sigma_2}=1$ ($\mathcal{C}_{\sigma_1\sigma_2}=-1$) indicates perfect correlation (anticorrelation) between the two particle momenta. The expectation values in Eq.~\eqref{eq_C} read
\begin{subequations}
\begin{align}
  \langle k_1 k_2 \rangle_{\sigma_1\sigma_2}&=\iint  \mathrm{d}k_1 \mathrm{d}k_2 \, k_1 k_2 \, \rho_2(k_1, \sigma_1 ; k_2, \sigma_2),\\        \langle k_1 \rangle_{\sigma_1}&=\int \mathrm{d}k_1\, k_1 \, \rho_1(k_1 \sigma_1),
  \end{align}
\end{subequations}
where $\rho_2(k_1, \sigma_1 ; k_2, \sigma_2)$ is the two-body density matrix in momentum space and $\rho_1(k_1,\sigma_1)$ is the one-body density matrix.

\par

In Fig.~\ref{2bdm}(a), we show the momentum correlation coefficient~\eqref{eq_C} for increasing pump strength for the degenerate case $\delta=0$ (the results do not change qualitatively for finite level splitting). 
Below threshold, particles with the same spin (black and grey curves in Fig.~\ref{2bdm}(a)) show perfectly anticorrelated momenta,  $\mathcal{C}_{\uparrow\uparrow}=\mathcal{C}_{\downarrow\downarrow}=-1$. However, as the threshold is surpassed by increasing the pump strength, the two-photon scattering with $\pm 2 k_c$ momentum transfer enhance the correlations between comoving particles with the same spin, eventually leading to uncorrelated momenta.

\par

By contrast, the momenta of particles in opposite spin states (blue curve in Fig.~\ref{2bdm}(a)) are uncorrelated below threshold, $\mathcal{C}_{\uparrow\downarrow}=\mathcal{C}_{\downarrow\uparrow}=0$. 
In the superradiant regime, however, the scattering processes from the pumps to the cavity induce unexpected positive correlations. These positive correlations saturate for intermediate pump strengths and vanish in the limit of very strong pump strenghts. 
\par

The behaviour of the correlation coefficient $\mathcal{C}_{\sigma_1\sigma_2}$ in Fig.~\ref{2bdm}(a) can be understood from the two-body density matrix in momentum space, shown in Figs.~\ref{2bdm}(b) and~\ref{2bdm}(c). There, the thermal background is visible as a smooth circle extending outside the Fermi surface below threshold ($\pm k_F$), indicated by the red lines. In the superradiant phase the scattering of photons by the atoms leads to the population of higher momentum states. Hence, the momentum distribution loses its sharpness and acquires tails at higher momenta which are responsible for the observed shape. In addition, forbidden states revealed by the dips along the diagonal, $k_2=k_1$, and the shifted diagonals, $k_2=k_1\pm 2 k_c$, appear in the two-body density matrix with the same spin, see Fig.~\ref{2bdm}(b). Likewise, for particles with opposite spins dips at $k_2=k_1\pm k_c$ develop (Fig.~\ref{2bdm}(c)).

\par

These diagonal dips can be understood as a consequence of the Pauli principle for fermions in a ring cavity. The Pauli principle forbids particles with the same spin to occupy the same momentum state, which explains the diagonal dip $k_2=k_1$ in Fig.~\ref{2bdm}(b). The off-diagonal dips are then a manifestation of the Pauli principle at higher momenta, indicating the absence of two-particle states that cannot be created via the interaction with the cavity photons because the required initial state is prohibited by the fermionic statistics. Since the state $\ket{k_1,\sigma_1;k_2=k_1,\sigma_2=\sigma_1}$ is forbidden by the Pauli principle, the Hamiltonian~\eqref{ham} cannot populate states with opposite spins and momenta $k_2=k_1\pm k_c$ or same spin and momenta $k_2=k_1\pm 2 k_c$. The propagation of the Pauli principle to higher momenta then forbids states with the same spin and momenta differing by even multiples of the cavity wave-vector, $\ket{ k_1, \sigma_1; k_2=k_1\pm 2jk_c,\sigma_2=\sigma_1}$ ($j\in\mathbb{N}_0$), and states with opposite spin and momenta differing by odd multiples of $k_c$, $\ket{ k_1, \sigma_1; k_2=k_1\pm (2j+1)k_c,\sigma_2=-\sigma_1}$. These higher-order dips become visible as the momentum distribution broadens in momentum space.

\par

For intermediate pump strengths above threshold the thermal background only moderately surpasses the $T=0$ Fermi surface. The forbidden states, i.e., the overlap of the off-diagonal dips with the thermal background, then mainly consist of counterpropagating pairs.
This leads to an excess of co-moving particles in the Fermi gas, which causes an increase of the momentum correlation coefficient~\eqref{eq_C} above threshold, both for particles having the same or opposite spin (Fig.~\ref{2bdm}(a)). However, the correlations increase faster for particles having opposite spin than for particles in the same spin state, since for the former the first off-diagonal forbidden states pertain to smaller momenta ($k_2=k_1\pm k_c$) than for the latter ($k_2=k_1\pm 2k_c$). With increasing pump strength, however, the momentum distribution broadens and the off-diagonal dips include a balanced contribution of both counterpropagating and co-moving pairs, leading to a decrease in correlation and eventually to uncorrelated particle momenta, $\mathcal{C}_{\uparrow\downarrow}=0$, far above threshold.

\section{\label{sec:level5}Conclusion and Outlook}
We explored the self-ordering of a fermionic gas coupled to the light fields of a tranversally-pumped ring resonator. The system is characterized by a continuous $U(1)$ translational symmetry and a disc rete $\mathbf{Z}_2$ spin inversion symmetry. The combined $U(1)\times \mathbf{Z}_2$ symmetry is spontaneously broken at the onset of a superradiant phase transition where the cavity modes become macroscopically populated. Above the transition threshold the atomic gas self-organizes in an state with antiferromagnetic character with spontaneously emerging density and spin waves. On a mean-field level, a very similar phase diagram could be realized for bosonic species. The effect of the quantum statistics is fundamental for low photon numbers where the mean-field approximation breaks down, which gives rise to new phases~\cite{fan2018magnetic}. Within our level of approximation the essential signature of the Fermi statistics is found in the two-body momentum correlations.
\par
In fact, cavity photons mediate strong cooperative effects between the atomic motion and the internal atomic dynamics. We accordingly observed strong correlations in momentum space: atoms in the same spin state show anticorrelation while atoms with opposite spin are characterized by unexpected positive correlations. Such correlations can be traced back to the propagation of the Pauli principle to higher momenta through the interaction with the cavity modes. Their nature 
is therefore a direct consequence of the fermionic statistics and can lead to the generation of strongly entangled states \cite{bergschneider2018correlations}. 
\par 
In conclusion, our system allows to explore a wealth of novel and interesting phenomena where the light fields are dynamically coupled to the atomic state. Such systems represent an optimal platform for the study of strongly correlated systems in many-body physics and condensed matter. In particular, the possibility of inducing a BCS-type pairing \cite{bardeen1957microscopic,bardeen1957theory,cooper1956bound} with cold atoms in optical cavities paves the way to the realization of light-induced superconductivity under controllable conditions and will be object of our future studies \cite{fausti2011light,demsar2016light,mitrano2016possible}. In  addition, interesting competing effects between the cavity wavelength and the density length-scale can be found at different filling factors, leading to the emergence of incommensurate spin and density structures. The role of inter-particle interactions, neglected in this work, has to be investigated as well \cite{sheikhan2018cavity}.  
\section*{Acknowledgments}
We thank Maria Luisa Chiofalo and Francesco Piazza for fruitful discussions. We acknowledge support by the Austrian Science Fund FWF through the Project SFB FoQuS F4013. W.\,N.\ acknowledges support from an ESQ fellowship of the Austrian Academy of Sciences (\"OAW). 
\providecommand{\newblock}{}

\widetext

 \appendix

 \section{\label{app:1}Derivation of the many-body Hamiltonian}
We consider a four-level fermionic atom coupled to two degenerate modes of a ring cavity, as represented in Fig.~\ref{scheme}. The atomic motion is restricted along the cavity axis ($z$-direction). The four states $\{\ket{\downarrow},\ket{\uparrow},\ket{e},\ket{e'}\}$ have energies $\{\hbar\omega_\downarrow=0,\hbar\omega_\uparrow,\hbar\omega_e,\hbar \omega_{e'}\}$, respectively. The atom is pumped from the side by two lasers with frequencies $\{\omega_{p1},\omega_{p2}\}$ and Rabi frequencies $\Omega_1$ and $\Omega_2$. The pumping lasers drive the transitions $\ket{\downarrow}\leftrightarrow\ket{e}$ and $\ket{\uparrow}\leftrightarrow\ket{e'}$, respectively. The cavity fields $\hat a_{\pm}$ couple both transitions $\ket{e'}\leftrightarrow\ket{\downarrow}$ and  $\ket{e}\leftrightarrow\ket{\uparrow}$. The single-particle Hamiltonian for this system is 
    \begin{align}
        H_1(t)&=\sum_{i\in\{\uparrow,e,e'\}}\hbar \omega_{i} \ket{i}\bra{i} +\hbar\omega_c\left(\hat a_+^\dagger \hat a_++\hat a_-^ \dagger \hat a_- \right)\nonumber\\ &\quad+ \hbar\left(\Omega_1 e^{i \omega_{p1} t}\ket{\downarrow}\bra{e} +\Omega_2e^{i \omega_{p2} t}\ket{\uparrow}\bra{e'}+\mathrm{h.c.}\right)\nonumber \\
&\quad+ \hbar g_0 \left( e^{i k_c z} \hat a_+ \ket{e'}\bra{\downarrow}+ e^{-i k_c z} \hat a_- \ket{e'}\bra{\downarrow}+ e^{i k_c z} \hat a_+ \ket{e}\bra{\uparrow}+ e^{-i k_c z} \hat a_- \ket{e}\bra{\uparrow}+\mathrm{h.c.}\right).
    \end{align} 
In order to eliminate the explicit time dependence we perform a unitary transformation to a frame where the lowest ground state $\ket{\downarrow}$ is at rest.  Applying the unitary transformation 
\begin{align}
U(t)&=\exp\left\{i\Big[\Big(\frac{\omega_{p1}+\omega_{p2}}{2}\Big)(\hat a_+^\dagger \hat a_+ +\hat a_-^\dagger \hat a_-)\Big]t\right\}\nonumber \\
&\quad \times \exp\left\{i\Big[\Big(\frac{\omega_{p1}+\omega_{p2}}{2}\Big)\ket{e'}\bra{e'} + \omega_{1} \ket{e}\bra{e}+\Big(\frac{\omega_{p2}-\omega_{p1}}{2}\Big)\ket{\uparrow}\bra{\uparrow} \Big]t\right\},
\end{align}

the time-independent Hamiltonian $\tilde H_1=U H_1 U^\dagger+i\hbar(\partial_t U) U^\dagger $ is
\begin{align}
        \tilde H_1&=\sum_{i\in\{\uparrow,e,e'\}}-\hbar \Delta_{i} \ket{i}\bra{i} -\hbar\Delta_c\left(\hat a_+^\dagger \hat a_++\hat a_-^ \dagger \hat a_- \right)+ \hbar\left(\Omega_1 \ket{\downarrow}\bra{e} +\Omega_2\ket{\uparrow}\bra{e'}+\mathrm{h.c.}\right)\nonumber \\
&\quad +\hbar g_0 \left( e^{i k_c z} \hat a_+ \ket{e'}\bra{\downarrow}+ e^{-i k_c z} \hat a_- \ket{e'}\bra{\downarrow}+ e^{i k_c z} \hat a_+ \ket{e}\bra{\uparrow}+e^{-i k_c z} \hat a_- \ket{e}\bra{\uparrow}+\mathrm{h.c.}\right),
    \end{align} 
where $\Delta_\downarrow=0$, $\Delta_\uparrow=({\omega_{p2}-\omega_{p1}})/{2}-\omega_\uparrow$, $\Delta_e=(\omega_{p1}+\omega_{p2})/2-\omega_e$, $\Delta_{e'}=\omega_{p1}-\omega_{e'}$ are the detunings of the four levels after the unitary transformation and $\Delta_c=({\omega_{p1}+\omega_{p2}})/{2}-\omega_\uparrow$ is the cavity detuning.
\par
For large atomic detunings, the excited states are adiabatically eliminated and the Hamiltonian reduces to the one of a spin-1/2 system, 
\begin{align}
        \tilde{H}&=\sum_{\sigma\in\{\downarrow,\uparrow\}}\left[\hbar \delta_{\sigma}+U_{0\sigma}\left(\hat a_+^\dagger \hat a_++\hat a_-^ \dagger \hat a_- +\hat a_+^\dagger \hat a_- e^{-2ik_cz}+\hat a_-^ \dagger \hat a_- e^{-2ik_cz}\right)\right] \ket{\sigma}\bra{\sigma}\nonumber \\
&\quad+ \hbar \eta\left(\ket{\uparrow}\bra{\downarrow}+\ket{\downarrow}\bra{\uparrow}\right)\left( e^{i k_c z} \hat a_+ + e^{-i k_c z} \hat a_-+\mathrm{h.c.}\right)
-\hbar\Delta_c\left(\hat a_+^\dagger \hat a_++\hat a_-^ \dagger \hat a_- \right),
    \end{align} 
    with $\delta_\downarrow=0$, $\delta_\uparrow=\omega_\uparrow-({\omega_{p2}-\omega_{p1}})/{2}$, $U_{0\uparrow}= \hbar g_0^2/\Delta_e$ and $U_{0\downarrow}=\hbar g_0^2/\Delta_{e'} $, and for the balanced case $\eta\equiv \Omega_1 g_0/\Delta_{e}=\Omega_2 g_0/\Delta_{e'}$.
    \par
    The many-body Hamiltonian for non-interacting particles is 
    \begin{equation}
        H=\int dz  \left(\hat \Psi^\dagger_\uparrow(z),\hat \Psi^\dagger_\downarrow(z)\right)
      \tilde H
      \begin{pmatrix}
      \hat \Psi_\uparrow(z) \\
      \hat \Psi_\downarrow(z)
     \end{pmatrix} ,   \end{equation}
 where $\hat\Psi_\sigma(z)$ are fermionic field operators fulfilling the anti-commutation relation $\{\hat\Psi_\sigma(z),\hat\Psi_{\sigma'}^\dagger(z')\}=\delta(z-z')\delta_{\sigma,\sigma'}$. Finally, by adding an external longitudinal magnetic field $\mathbf{B}=(0,0,B_z)$ we obtain Hamiltonian \eqref{ham} in the main text.
 \section{\label{app:2}Heisenberg equations of motion for the cavity fields}
 The Heisenberg equations of motion for the photonic field operators are
 \begin{subequations}
\begin{align}
    i\hbar \partial_t \hat a_+&=[\hat a_+,H]=-\hbar(\Delta_c-U_0N+i\kappa)\hat a_++U_0 \int dz e^{2ik_cz}\hat n(z) \hat a_-+\hbar \eta \int dz e^{ik_cz }\hat S_x(z),\\
    i\hbar \partial_t \hat a_-&=[\hat a_-,H]=-\hbar(\Delta_c-U_0N+i\kappa)\hat a_-+U_0 \int dz e^{-2ik_cz}\hat n(z) \hat a_++\hbar \eta \int dz e^{-ik_cz }\hat S_x(z).
\end{align}
\end{subequations}
Here $\hat n(z)=\hat \Psi_\uparrow^\dagger(z)\hat \Psi_\uparrow(z)+\hat \Psi_\downarrow^\dagger(z)\hat \Psi_\downarrow(z)$ is the atomic density operator and $\hat S_x(z)=[\hat \Psi_\uparrow^\dagger(z)\hat \Psi_\downarrow(z)+\hat \Psi_\downarrow^\dagger(z)\hat \Psi_\uparrow(z)]/2$ is the local spin operator in the $x$-direction. When the photonic operators evolve on a faster timescale with respect to the atomic dynamics, we can consider the stationary value of the photonic operators and express the mean-field average $\langle \hat a_\pm\rangle =\alpha_\pm$ as a function of mean-field atomic averages. The former equations become
 \begin{subequations}
\begin{align}
    -\hbar(\Delta_c-U_0N+i\kappa) \alpha_++U_0 \mathcal{N}_{2k_c}  \alpha_-+\hbar \eta \Theta&=0,\\
  -\hbar(\Delta_c-U_0N+i\kappa) \alpha_-+U_0 \mathcal{N}_{2k_c}^* \alpha_++\hbar \eta \Theta^*&=0.
\end{align}
  \label{eqmean}
\end{subequations}
Here, $\mathcal{N}_{2k_c}$ and $\Theta$ are the atomic averages defined in the main text in Eq.~\eqref{cdw} and Eq.~\eqref{sdw}, respectively. The spin order parameter $\Theta$ plays the fundamental role in the superradiant phase transition. Note that a macroscopic cavity field can only be induced by the emergence of a spin self-ordered state (non-vanishing $\Theta$). In contrast, the density self-ordering (non-vanishing $\mathcal{N}_{2k_c}$) does not act as a source of cavity photons in Eqs.~\eqref{eqmean} but only induces the mixing between the two modes $\alpha_\pm$. By solving the coupled Eqs.~\eqref{eqmean} we obtain Eqs.~\eqref{eqalphas} in the main text.

 \section{\label{app:3}Dependence of the transition threshold on the filling factor}

\begin{figure}[h]
\centering
\includegraphics[width=0.6\textwidth]{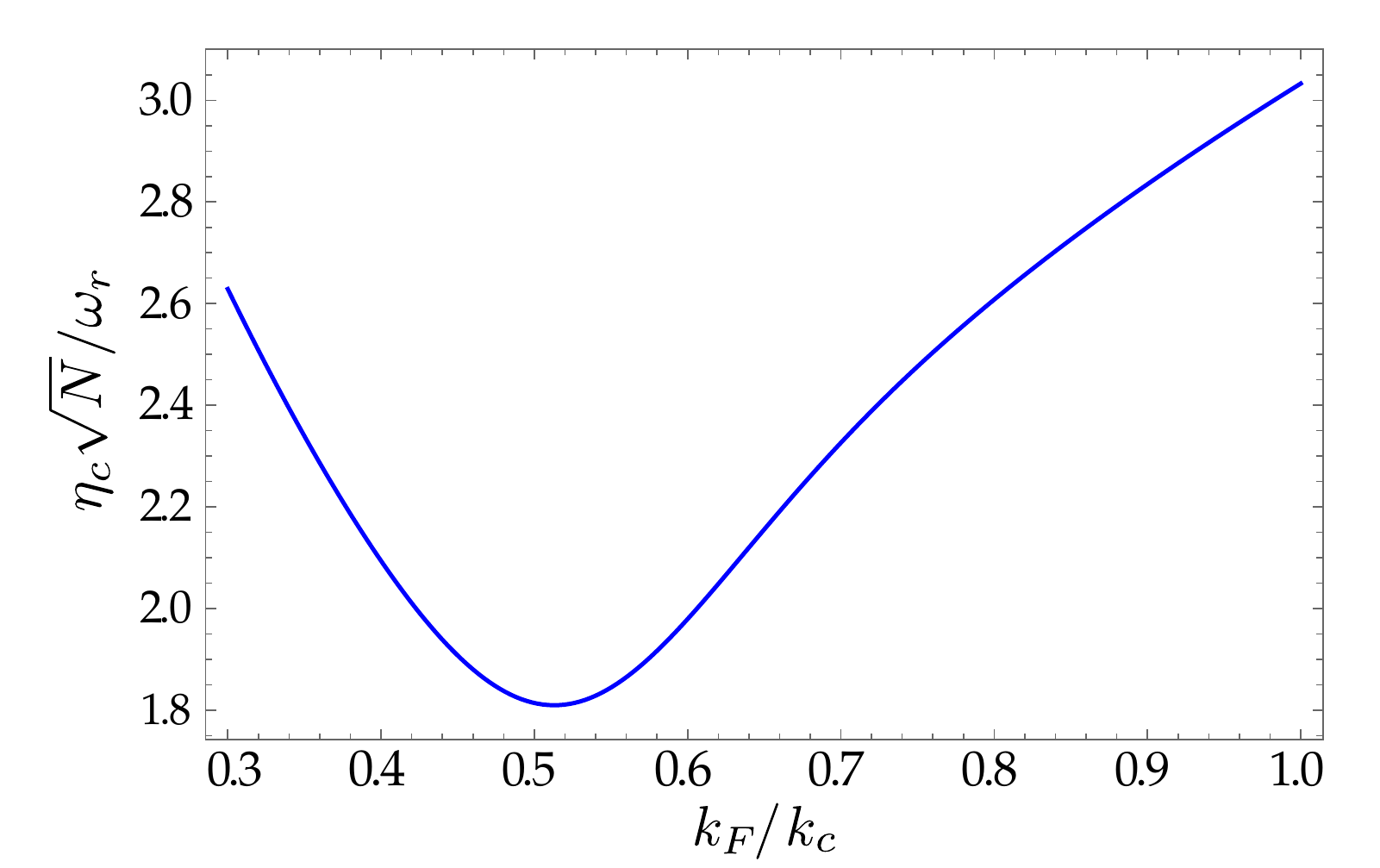}
\caption{Critical threshold at fixed temperature for different filling factors, as obtained from Eq.~\eqref{critpump}.}
\label{fig:fillingfactor}
\end{figure}

The analysis presented in the main text was performed for half filling $k_F/k_c=1/2$. Here we would like to show the effect of different filling on the phases presented above. This can most easily be done by looking at the dependence of the critical pump strength on the filling factor (see~\ref{fig:fillingfactor}). For finite temperature the threshold is suppressed at half-filling due to the nesting condition, as it was already mentioned in the main text. The presence of harmonic confinement introduces inhomogeneities,  influencing the atomic susceptibility as already noticed in~\cite{keeling2014fermionic}. Despite harmonic confinement, commensurate effects are still visible in the critical pump-cavity detuning curve when the nesting condition holds in the trap center. From~\ref{fig:fillingfactor} it can be seen that despite the fact that the critical pump strength is higher at filling factors different from half-filling, the system will still undergo a superradiant phase transition at threshold. Therefore, the predicted phases can in general be observed for different filling factors. Similar characteristics were found in single component Fermi gases for both red and blue detuning with respect to the atomic transition frequency~\cite{piazza2014umklapp,mivehvar2017superradiant}.

\end{document}